\documentclass[12pt]{article}
\usepackage{amsmath}
\usepackage{amssymb}
\usepackage{amsfonts}
\usepackage{graphics}

\numberwithin{equation}{section}
\newcommand{\field}[1]{\mathbb{#1}}
\newcommand{\R}{\field{R}}

\title{
Quasi-exactly solvable symmetrized quartic and sextic polynomial oscillators}
\author{C. Quesne\\
{\small \sl Physique Nucl\'eaire Th\'eorique et Physique Math\'ematique,}\\ 
{\small \sl Universit\'e Libre de Bruxelles, Campus de la Plaine CP229,} \\ 
{\small \sl Boulevard~du Triomphe, B-1050 Brussels, Belgium} \\
{\small \sl cquesne@ulb.ac.be}}
\date{ }
\begin{document}
\baselineskip=22pt plus 1pt minus 1pt
\maketitle

\begin{abstract} 
The symmetrized quartic polynomial oscillator is shown to admit an sl(2,$\R$) algebraization. Some simple quasi-exactly solvable (QES) solutions are exhibited. A new symmetrized sextic polynomial oscillator is introduced and proved to be QES by explicitly deriving some exact, closed-form solutions by resorting to the functional Bethe ansatz method. Such polynomial oscillators include two categories of QES potentials: the first one containing the well-known analytic sextic potentials as a subset, and the second one of novel potentials with no counterpart in such a class.
\end{abstract}

\noindent
Short title: Quasi-exactly solvable symmetrized polynomial oscillators

\noindent
PACS Nos.: 03.65.Fd, 03.65.Ge
%
%
\newpage
\section{Introduction}

Exact solutions of the Schr\"odinger equation are very useful for practical physical problems because they may provide a good starting point for developing a constructive perturbation theory or for suggesting trial functions in variational calculus applicable to some more realistic equations appearing in those problems.\par
%
%
A first category of exact solutions pertains to the so-called exactly solvable (ES) Schr\"odinger equations, \textit{i.e.}, those equations for which all the eigenstates can be algebraically determined. This category splits into several subcategories, among which we may quote those containing piece-wise constant point interactions (see, \textit{e.g.}, the rectangular potential hole in \cite{flugge}) and those defined in terms of a polynomially solvable analytic potential (see, \textit{e.g.}, one of their lists in \cite{cooper} and the recent development of the exceptional orthogonal polynomials in \cite{gomez09, gomez10, cq, odake09, odake11}). For the latter, the eigenfunctions are constructed in closed form by using the theory of special functions (as well as Darboux transformations), while for the former the solutions are obtained by matching the wavefunctions and their first derivatives at the non-analyticity points of the potential.\par
%
%
A second category of exact solutions belongs to the so-called quasi-exactly solvable (QES) Schr\"odinger equations \cite{leach, turbiner87, turbiner88, ushveridze, gonzalez, turbiner16, ho, zhang, agboola12, agboola13, agboola14, ciftci} . These occupy an intermediate place between ES and non-solvable ones in the sense that only a finite number of eigenstates can be found explicitly by algebraic means while the remaining ones remain unknown. Several procedures are used to construct them, among which the two easiest ones are those based on an sl(2, $\R$) algebraization \cite{turbiner87, turbiner88, ushveridze, gonzalez, turbiner16} or on the functional Bethe ansatz method \cite{ho, zhang, agboola12, agboola13, agboola14}. Another approach relies on the recursion relation method (see, \textit{e.g.}, \cite{ciftci}).\par
%
%
It was recently suggested \cite{znojil16a, sasaki16a, znojil16b, sasaki16b, sasaki16c} that, apart from those above-mentioned interaction models, the ES status should also be attributed to less common ones, for which the real line of coordinates is splitted into subintervals wherein the potential admits different definitions, while being continuous on the whole line, and the wavefunctions remain piece-wise proportional to special functions, while being matched, as well as their first derivatives, at the subinterval limit points.\par
%
%
A similar extension has also been proposed in the case of QES Schr\"odinger equations. In a recent work \cite{znojil16c}, Znojil indeed introduced an analyticity-violating symmetrization of the quartic polynomial oscillator. Such a symmetrization amounts to part the real line into two subintervals $\R^-$ and $\R^+$, on which the corresponding Schr\"odinger equations are solved by using the recursion relation approach, and to match the resulting wavefunctions and their first derivatives at $x=0$. The Schr\"odinger equation on the whole line then turns out to be QES in an extended sense. As a consequence, it may be seen as completing the list of (analytic) QES anharmonic oscillators, which contains sextic polynomial oscillators \cite{leach, turbiner87}, but no quartic one.\par  
%
%
After commenting in sec.~2 on the QES symmetrized quartic polynomial oscillator of Ref.~\cite{znojil16c}, we will turn ourselves in sec.~3 to the main purpose of the present work, namely proposing a new QES symmetrized sextic polynomial oscillator and deriving some explicit examples of it. Finally, sec.~4 will contain some concluding remarks.\par
%
%
\section{QES symmetrized quartic polynomial oscillator}

As in \cite{znojil16c}, let us consider the symmetrized quartic polynomial oscillator
\begin{equation}
  V(x) = x^4 - s |x|^3 + r x^2 - q |x| = 
  \begin{cases}
    x^4 + s x^3 + r x^2 + q x & \text{if $x<0$}, \\
    x^4 - s x^3 + r x^2 - q x & \text{if $x>0$},
  \end{cases} 
\end{equation}
which is non-analytic and defined separately on the two semi-infinite intervals $x<0$ and $x>0$. Here $q$, $r$, and $s$ are three real constants. The corresponding Schr\" odinger equation
\begin{equation}
  \left(- \frac{d^2}{dx^2} + V(x)\right) \psi(x) = E \psi(x)  \label{eq:SE}
\end{equation}
may be considered separately on these two intervals. Physically acceptable wavefunctions should tend to zero at infinity and be continuous everywhere, as well as their first derivative. Since the potential is symmetric under parity, they must also be either symmetric or antisymmetric. We may therefore solve eq.~(\ref{eq:SE}) in one of the intervals, \textit{e.g.}, $x<0$, then impose that in the other interval $x>0$, $\psi(x) = \epsilon \psi(-x)$ with $\epsilon = +1$ or $\epsilon = -1$ according to whether $\psi$ is even or odd.\footnote{One may alternatively say that there are two types of eigenfunctions, one satisfying Neumann boundary condition $\psi'(0) = 0$ at $x=0$ (for $\epsilon=+1$) and the other the Dirichlet condition $\psi(0)=0$ (for $\epsilon=-1$).}\par
%
%
QES solutions of (\ref{eq:SE}) may be obtained by assuming that $\psi(x)$ takes the form
\begin{equation}
  \psi_n^{(\epsilon)}(x) = e^{W(x)} \phi_n^{(\epsilon)}(x), \qquad W(x) = - \tfrac{1}{3} |x|^3 + a x^2 - b |x|,
  \label{eq:QES-4}
\end{equation}
with
\begin{equation}
  \phi_n^{(\epsilon)} = 
    \begin{cases}
       \sum_{k=0}^n v_k x^k & \text{if $x<0$}, \\
       \epsilon \sum_{k=0}^n (-1)^k v_k x^k & \text{if $x>0$}.
   \end{cases}  \label{eq:phi}
\end{equation}
Here $a$, $b$ are two real constants, and $v_k$ ($k=0$, 1, \ldots, $n$) are some expansion coefficients to be determined. Such functions have the right behaviour for $x \to \pm \infty$. On the other hand, the continuity of $\psi_n^{(\epsilon)}(x)$ and of its derivative at the origin imposes the additional condition
\begin{equation}
\begin{split}
  & v_1 + b v_0 = 0 \qquad \text{if $\epsilon = +1$}, \\
  & v_0 = 0 \qquad \text{if $\epsilon = -1$}. 
\end{split}  \label{eq:constraint-4}
\end{equation}
\par
%
%
Inserting eq.~(\ref{eq:QES-4}) in (\ref{eq:SE}) and setting
\begin{equation}
  s = 4a \qquad \text{and} \qquad r = 4a^2 + 2b
\end{equation}
in order to eliminate some high-order terms, we arrive at the differential equation
\begin{equation}
  \left( - \frac{d^2}{dx^2} - 2(x^2 + 2ax + b)\frac{d}{dx} + (q - 4ab - 2)x - b^2 - 2a\right) \phi_n(x) =
  E_n \phi_n(x)  \label{eq:heun}
\end{equation}
on the interval $x<0$. This is a (triconfluent) Heun equation \cite{ronveaux}. Hence, provided we also set
\begin{equation}
  q = 4ab + 2n + 2,
\end{equation}
the second-order differential operator $h$ on its left-hand side can be written as a second-degree polynomial \cite{turbiner16}
\begin{equation}
  h = - J^-_n J^-_n - 2 J^+_n - 4a J^0_n - 2b J^-_n - 2(n+1)a - b^2  \label{eq:J}
\end{equation}
in the sl(2,$\R$) generators
\begin{equation}
  J^+_n = x^2 \frac{d}{dx} - nx, \qquad J^0_n = x \frac{d}{dx} - \frac{n}{2}, \qquad J^-_n = \frac{d}{dx}
\end{equation}
acting in the ($n+1$)-dimensional representation in the space of polynomials of order not higher than $n$, ${\cal P}_{n+1} = \langle 1, x, x^2, \ldots, x^n\rangle$. On using the relations $J^+_n x^k = (k-n) x^{k+1}$, $J^0_n x^k = (k - \frac{n}{2}) x^k$, $J^-_n x^k = k x^{k-1}$, it is straightforward to explicitly write the matrix of $h$ in the space ${\cal P}_{n+1}$. Its diagonalization, followed by the imposition of constraint (\ref{eq:constraint-4}), will then provide the admissible energies $E_n^{(\epsilon)}$ and wavefunctions $\psi_n^{(\epsilon)}(x)$.\par
%
%
In \cite{znojil16c}, Znojil solved the rather complicated $n=2$, $\epsilon=+1$ case by the recursion relation method. In the remainder of this section, we plan to exhibit some simpler examples, which can be directly derived by using (\ref{eq:heun})--(\ref{eq:J}).\par
%
%
{}For $n=0$, we obtain $E_0 = - 2a - b^2$ and $\phi_0(x) = 1$ on $x<0$. Since $v_0=1$ and $v_1=0$, according to (\ref{eq:constraint-4}), the resulting wavefunction on the whole real line cannot be odd, but it can be even if the condition $b=0$ is satisfied. We conclude that the potential
\begin{equation}
  V(x) = x^4 - 4a |x|^3 + 4a^2 x^2 - 2 |x|
\end{equation}
has an energy eigenvalue $E_0^{(+)}$ and associated wavefunction $\psi_0^{(+)}(x)$, given by
\begin{equation}
  E_0^{(+)} = - 2a, \qquad \psi_0^{(+)}(x) = e^{-\frac{1}{3} |x|^3 + a x^2},
\end{equation}
corresponding to a ground state.\par
%
%
Turning now ourselves to the $n=1$ case, we arrive at a second-degree equation for $E_1$, whose solutions $E_{1\pm} = - 4a - b^2 \pm 2\sqrt{a^2 - b}$ are real provided $a^2 \ge b$, which imposes $b \le 0$ or $b > 0$ and either $a \ge \sqrt{b}$ or $a \le - \sqrt{b}$. The corresponding eigenfunctions on $x<0$ are $\phi_{1\pm}(x) = x+a \pm \sqrt{a^2 - b}$. Hence $v_0 = a \pm \sqrt{a^2 - b}$ and $v_1 = 1$. On the whole line, we can get an odd wavefunction if $a = \mp \sqrt{a^2 - b}$, which imposes $b=0$ and $a<0$ (resp.\ $a>0$) for $\phi_{1+}$ (resp.\ $\phi_{1-}$). We conclude that the potential
\begin{equation}
  V(x) = x^4 - 4a |x|^3 + 4a^2 x^2 - 4 |x|
\end{equation}
has an energy eigenvalue $E_1^{(-)}$ and associated wavefunction $\psi_1^{(-)}(x)$, given by
\begin{equation}
  E_1^{(-)} = - 6a, \qquad \psi_1^{(-)}(x) = e^{- \frac{1}{3} |x|^3 + ax^2} x,
\end{equation}
corresponding to a first-excited state.\par
%
%
We may also ask whether an even wavefunction can be obtained for $n=1$. From (\ref{eq:constraint-4}), we directly get that $b \ne 0$ and $a \pm \sqrt{a^2 - b} = - 1/b$, where the upper (resp.\ lower) sign corresponds to $\phi_{1+}$ (resp.\ $\phi_{1-}$). This implies that $a$ can be expressed in terms of $b$ as $a = -(b^3+1)/(2b)$. As a result, $\sqrt{a^2 - b} = (b^3-1)/(2b)$ and $a + \sqrt{a^2 - b} = -1/b$ if $b<0$ or $b>1$, whereas $\sqrt{a^2 - b} = (1-b^3)/(2b)$ and $a - \sqrt{a^2 - b} = -1/b$ if $0<b\le 1$. In both cases, $E_{1+} = E_{1-} = (2b^3+1)/b$. Hence, the potential  
\begin{equation}
  V(x) = x^4 + \frac{2(b^3+1)}{b} |x|^3 + \frac{b^6 + 4b^3 +1}{b^2} x^2 + 2(b^3-1) |x|, \qquad b\ne 0,
\end{equation}
has an energy eigenvalue $E_1^{(+)}$ and associated wavefunction $\psi_1^{(+)}(x)$, given by
\begin{equation}
  E_1^{(+)} = \frac{2b^3+1}{b}, \qquad \psi_1^{(+)}(x) = e^{- \frac{1}{3} |x|^3 - \frac{b^3+1}{2b} x^2
  - b |x|} \left(|x| + \frac{1}{b}\right),
\end{equation}
corresponding to a ground state for $b>0$ and to a second-excited state for $b<0$.\par
%
%
\section{QES symmetrized sextic polynomial oscillator}
\setcounter{equation}{0}

\begin{figure}[h]
\begin{center}
\includegraphics{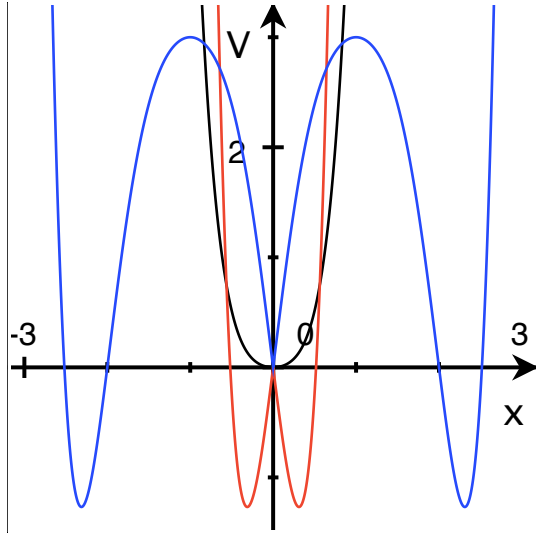}
\caption{The potential $V(x)$, defined in eq.~(\ref{eq:V-0+}), in terms of $x$ for $a=0$ (black line), $a=+1$ (red line), and $a=-1$ (blue line). In all cases, $b=-1$ and $E_0^{(+)}=2$.}
\end{center}
\end{figure}
\par
%
%
\begin{figure}[h]
\begin{center}
\includegraphics{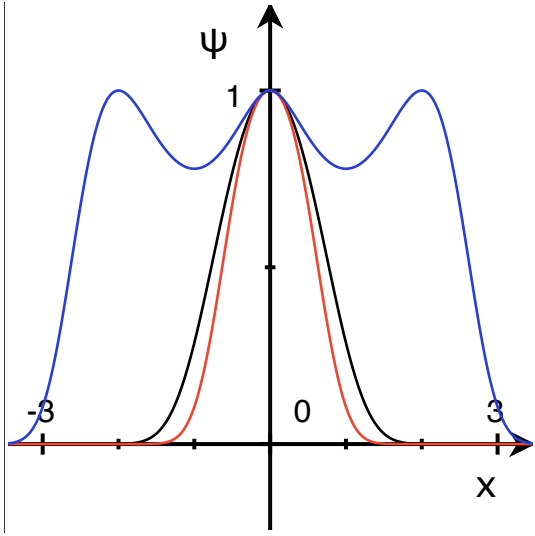}
\caption{The QES ground-state wavefunction $\psi_0^{(+)}(x)$, defined in eq.~(\ref{eq:psi-0+}), in terms of $x$ for $a=0$ (black line), $a=+1$ (red line), and $a=-1$ (blue line). In all cases, $b=-1$ and $E_0^{(+)}=2$.}
\end{center}
\end{figure}
\par
%
%
\begin{figure}[h]
\begin{center}
\includegraphics{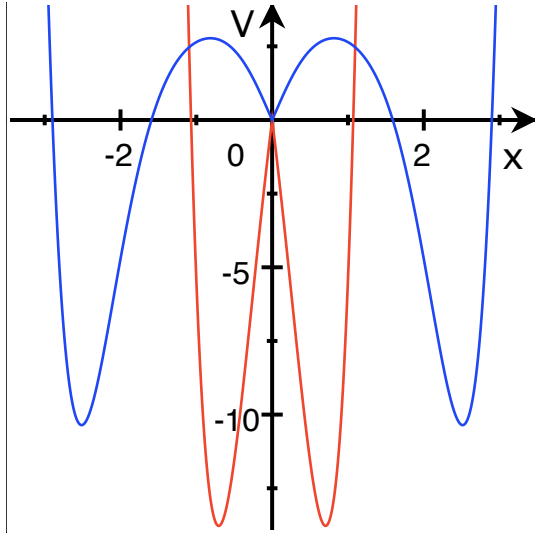}
\caption{The potential $V(x)$, defined in eq.~(\ref{eq:V-1+}), in terms of $x$ for $a=+1$, $b=-1$, $c=-1.949788$ (red line), and $a=-1$, $b=-1$, $c=-0.458984$ (blue line). The corresponding energy is $E_1^{(+)}= 5.80187$ and $E_1^{(+)}=2.21067$, respectively.}
\end{center}
\end{figure}
\par
%
%
\begin{figure}[h]
\begin{center}
\includegraphics{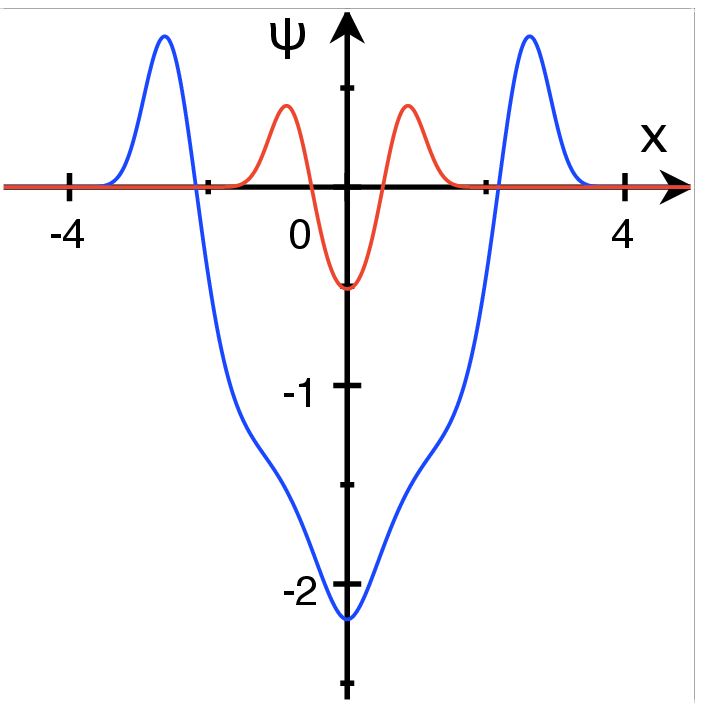}
\caption{The QES second-excited state wavefunction $\psi_1^{(+)}(x)$, defined in eq.~(\ref{eq:psi-1+}), in terms of $x$ for $a=+1$, $b=-1$, $c=-1.949788$ (red line), and $a=-1$, $b=-1$, $c=-0.458984$ (blue line). The corresponding energy is $E_1^{(+)}= 5.80187$ and $E_1^{(+)}=2.21067$, respectively.}
\end{center}
\end{figure}
\par
%
%
Let us now consider the symmetrized sextic polynomial oscillator
\begin{equation}
  V(x) = x^6 - u |x|^5 + t x^4 - s |x|^3 + r x^2 - q |x|,  \label{eq:sextic}
\end{equation}
where $q$, $r$, $s$, $t$, and $u$ are five real constants, and let us proceed as in sec.~2. QES solutions of (\ref{eq:SE}) with potential (\ref{eq:sextic}) may be obtained by assuming that
\begin{equation}
  \psi_n^{(\epsilon)}(x) = e^{W(x)} \phi_n^{(\epsilon)}(x), \qquad W(x) = - \frac{1}{4} x^4 - a |x|^3 + b x^2
  - c |x|,
\end{equation}
where $a$, $b$, $c$ are three real constants and $\phi_n^{(\epsilon)}(x)$ is still given by eq.~(\ref{eq:phi}), but with eq.~(\ref{eq:constraint-4}) replaced by the constraint
\begin{equation}
\begin{split}
  & v_1 + c v_0 = 0 \qquad \text{if $\epsilon = +1$}, \\
  & v_0 = 0 \qquad \text{if $\epsilon = -1$}. 
\end{split}  \label{eq:constraint-6}
\end{equation}
\par
%
%
By setting now
\begin{equation}
  u = - 6a, \qquad t = 9a^2 - 4b, \qquad s = 12ab - 2c,
\end{equation}
we get on $x<0$ the differential equation
\begin{align}
  &\biggl(- \frac{d^2}{dx^2} + 2(x^3 - 3ax^2 - 2bx - c) \frac{d}{dx} + (r - 4b^2 - 6ac + 3)x^2 
       + (q -4bc - 6a)x \nonumber \\
  & \quad -c^2 - 2b\biggr) \phi_n(x) = E_n \phi_n(x),
\end{align}
which is (a special case of) a generalized Heun equation. Its $n$th-degree polynomial solutions $\phi_n(x) = \prod_{i=1}^n (x-x_i)$ with distinct roots $x_1$, $x_2$, \ldots, $x_n$, can be found by applying the functional Bethe ansatz method. Direct application of theorem 1.1 of ref.~\cite{zhang} leads to the following conditions
\begin{equation}
\begin{split}
  & r = 4b^2 + 6ac - 2n - 3, \qquad q = -2 \sum_{i=1}^n x_i + 6a(n+1) + 4bc, \\
  & E = 2 \sum_{i=1}^n x_i^2 - 6a \sum_{i=1}^n x_i - 2b(2n+1) - c^2,
\end{split}
\end{equation}
where the roots $x_1$, $x_2$, \ldots, $x_n$ satisfy the Bethe ansatz equations
\begin{equation}
  \sum_{\substack{
    j=1 \\
    j\ne i}} ^n \frac{1}{x_i - x_j} - x_i^3 + 3a x_i^2 + 2b x_i + c = 0, \qquad i=1, 2, \ldots, n.
\end{equation}
On the resulting $\phi_n(x)$, it then remains to impose constraint (\ref{eq:constraint-6}).\par
%
%
The $n=0$ case is straightforwardly solved. As for the quartic oscillator, there is no odd eigenfunction, but there is an even one provided $c=0$. Hence, the potential
\begin{equation}
  V(x) = x^6 + 6a |x|^5 + (9a^2 - 4b) x^4 - 12ab |x|^3 + (4b^2 - 3) x^2 - 6a |x|  \label{eq:V-0+}
\end{equation}
has an energy eigenvalue $E_0^{(+)}$ and associated wavefunction $\psi_0^{(+)}(x)$, given by
\begin{equation}
  E_0^{(+)} = - 2b, \qquad \psi_0^{(+)}(x) = e^{- \frac{1}{4} x^4 - a |x|^3 + b x^2},  \label{eq:psi-0+}
\end{equation}
corresponding to a ground state.\par
%
%
{}For $n=1$, there is a single Bethe ansatz equation to be solved, namely $x_1^3 - 3a x_1^2 - 2b x_1 - c = 0$. Since $v_0 = -x_1$ and $v_1=1$, we can get an even or odd wavefucntion by appropriately choosing the parameters. The function is odd if $x_1=0$, implying $c=0$. Hence the potential
\begin{equation}
  V(x) = x^6 + 6a |x|^5 + (9a^2 - 4b) x^4 - 12ab |x|^3 + (4b^2 - 5) x^2 - 12a |x|
\end{equation}
has an energy eigenvalue $E_1^{(-)}$ and associated wavefunction $\psi_i^{(-)}(x)$, given by
\begin{equation}
  E_1^{(-)} = - 6b, \qquad \psi_1^{(-)}(x) = e^{- \frac{1}{4} x^4 - a |x|^3 + b x^2} x,  \label{eq:E-1-}
\end{equation}
corresponding to a first-excited state. On the other hand, the function is even if $x_1 = 1/c$, where $c$ must be a nonvanishing solution of the quartic equation $c^4 + 2b c^2 + 3a c - 1=0$. Then, for such a $c$ value, the potential
\begin{align}
  V(x) & = x^6 + 6a |x|^5 + (9a^2 - 4b) x^4 - (12ab - 2c) |x|^3 + (4b^2 + 6ac - 5) x^2 \nonumber \\
  & \quad - \left(12a + 4bc - \frac{2}{c}\right) |x|  \label{eq:V-1+}
\end{align}
admits $E_1^{(+)}$ and $\psi_1^{(+)}(x)$, given by
\begin{equation}
  E_1^{(+)} = - 6b - c^2 - \frac{6a}{c} + \frac{2}{c^2}, \qquad \psi_1^{(+)}(x) = e^{- \frac{1}{4} x^4 
  - a |x|^3 + b x^2 - c |x|} \left(|x| + \frac{1}{c}\right),  \label{eq:psi-1+}
\end{equation}
representing a ground state for $c>0$ and a second-excited state for $c<0$.\par
%
%
{}For $n=2$ and $\epsilon=-1$, we get $x_1=0$ and $x_2=1/c$, where $c$ is a nonvanishing solution of $2c^4 + 2b c^2 + 3a c - 1=0$. This yields
\begin{equation}
\begin{split}
  V(x) & = x^6 + 6a |x|^5 + (9a^2 - 4b) x^4 - (12ab - 2c) |x|^3 + (4b^2 + 6ac - 7) x^2 \\
  &  \quad - \left(18a + 4bc - \frac{2}{c}\right) |x|, \\
  E_2^{(-)} &= - 10b - c^2 - \frac{6a}{c} + \frac{2}{c^2}, \\
  \psi_2^{(-)}(x) &= e^{- \frac{1}{4} x^4 - a |x|^3 + b x^2 - c |x|} x \left(|x| + \frac{1}{c}\right),
\end{split}
\end{equation}
where $\psi_2^{(-)}(x)$ corresponds to a first-excited (resp.\ third-excited) state for $c>0$ (resp.\ $c<0$).\par
%
%
{}Furthermore, for $n=2$ and $\epsilon=+1$, we obtain
\begin{equation}
\begin{split}
  V(x) & = x^6 + 6a |x|^5 + (9a^2 - 4b) x^4 - (12ab - 2c) |x|^3 + (4b^2 + 6ac - 7) x^2 \\
  &  \quad - (18a + 4bc - 2x_1 - 2x_2) |x|, \\
  E_2^{(+)} &= - 10b - c^2 - 6a(x_1+x_2) + 2(x_1^2+x_2^2), \\ 
  \psi_2^{(-)}(x) &= e^{- \frac{1}{4} x^4 - a |x|^3 + b x^2 - c |x|} (|x|+x_1) (|x|+x_2),
\end{split}  \label{eq:V-2+}
\end{equation}
where $x_1$ and $x_2$ are solutions different from zero and $1/c$ of the set of equations
\begin{equation}
  x_1^3 - 3a x_1^2 - 2b x_1 - c = - (x_2^3 - 3a x_2^2 - 2b x_2 - c) = \frac{1}{x_1-x_2}, \qquad
  x_1 + x_2 = cx_1x_2.  \label{eq:bethe}
\end{equation}
\par
%
%
The set of QES symmetrized sextic polynomial oscillators that we have constructed here can be separated into two categories.
\begin{itemize}
\item[(i)] The symmetrized sextic oscillators whose known wavefunction is of `natural' parity, i.e., such that $\epsilon = (-1)^n$. These include those associated with the cases $(n,\epsilon) = (0,+1)$, $(1,-1)$, and $(2,+1)$, considered above. It is straightforward to see that they comprise as special cases the well-known (analytic) sextic oscillators of ref.~\cite{turbiner87}. This is obvious if we set $a=0$ in eqs.~(\ref{eq:V-0+})--(\ref{eq:E-1-}), corresponding to $(n,\epsilon) = (0,+1)$ and $(1,-1)$. In the case of eq.~(\ref{eq:V-2+}) for $(n,\epsilon) = (2,+1)$, on assuming $a = c = x_1+x_2 = 0$, the set of equations (\ref{eq:bethe}) reduces to a single equation $2x_1^4 - 4b x_1^2 - 1=0$, leading to $(x_1^2)_{\pm} = \frac{1}{2}(2b \pm \sqrt{4b^2+2})$. As a consequence, $E_{2\pm}^{(+)} = - 6b \pm 2\sqrt{4b^2+2}$ and $\psi_{2\pm}^{(+)}(x) = \exp(- \frac{1}{4}x^4 + bx^2) (x^2 - b \mp \frac{1}{2} \sqrt{4b^2+2})$, in agreement with \cite{turbiner87}.  
\item[(ii)] The symmetrized sextic oscillators whose known wavefunction is of `unnatural' parity, i.e., such that $\epsilon = (-1)^{n+1}$. These include those associated with the cases $(n,\epsilon) = (1,+1)$ and $(2,-1)$, considered above. Such symmetrized oscillators do not comprise any analytic sextic oscillator as found in ref.~\cite{turbiner87}.
\end{itemize}
\par
%
%
In figs.~1 and 2, some examples of potential and associated wavefunction are displayed for the case $(n,\epsilon) = (0,+1)$, belonging to the first category. They are compared with the corresponding results for the analytic potential with $a=0$. In figs.~3 and 4, some examples of potential and associated wavefunction are plotted for the case $(n,\epsilon) = (1,+1)$, belonging to the second category.\par 
%
%
\section{Conclusion}

In the present work, we have shown that the previously introduced QES symmetrized quartic polynomial oscillator  admits an sl(2,$\R$) algebraization and we have used the latter to provide some simple examples of bound-state wavefunctions at certain couplings and energies.\par
%
%
More importantly, we have then proposed a new symmetrized sextic polynomial oscillator and we have proved its QES nature by deriving some exact, closed-form solutions of the corresponding Schr\"odinger equation by resorting to the functional Bethe ansatz method. Furthermore, we have shown that such symmetrized sextic oscillators can be separated into two categories: a first set generalizing the well-known analytic sextic oscillators of ref.~\cite{turbiner87}, which form a subset, and a second set of novel QES potentials with no counterpart in that class of analytic sextic oscillators.\par
%
%
The present work enlarges the class of polynomial oscillators for which some exact solutions are available and may serve as a first approximation in the numerous realistic physical problems wherein anharmonicity effects are present, such as in molecular physics.\par
%
%
Trying to extend other known QES analytic potentials by allowing some additional analyticity-violating terms would provide an interesting topic for future investigation.\par
%
%

\end{document}